\documentclass{sf2a-conf2016}
\usepackage{graphicx}
\usepackage{hyperref}
\usepackage[]{natbib}  
\usepackage{epstopdf}
\usepackage{floatrow}
\usepackage{txfonts}

\def\BibTeX{{\rm B\kern-.05em{\sc i\kern-.025em b}\kern-.08em
    T\kern-.1667em\lower.7ex\hbox{E}\kern-.125emX}}
\bibpunct{(}{)}{;}{a}{}{,}  

\begin{document}

\TitreGlobal{SF2A 2016}


\title{Magnetic activity of seismic solar analogs}

\runningtitle{Magnetic activity of seismic solar analogs}

\author{D. Salabert}\address{Laboratoire AIM, CEA/DRF-CNRS, Universit\'e Paris 7 Diderot, IRFU/SAp, Centre de Saclay, 91191 Gif-sur-Yvette, France}

\author{R.~A. Garc\'ia$^1$}

\author{P.~G. Beck$^1$}

\author{the SSA team}

\setcounter{page}{237}


\maketitle


\begin{abstract}
We present our latest results on the solar--stellar connection by studying 18 solar analogs that we identified among the {\it Kepler} seismic sample \citep{salabert16a}. We measured their magnetic activity properties using observations collected by the {\it Kepler} satellite and the ground-based, high-resolution \textsc{Hermes} spectrograph. The photospheric (S${_\textrm{ph}}$) and chromospheric ($\mathcal{S}$) magnetic activity proxies of these seismic solar analogs are compared in relation to solar activity. We show that the activity of the Sun is actually comparable to the activity of the seismic solar analogs. Furthermore, we report on the discovery of temporal variability in the acoustic frequencies of the young (1\,Gyr-old) solar analog KIC\,10644253 with a modulation of about 1.5 years, which agrees with the derived photospheric activity \citep{salabert16b}. It could actually be the signature of the short-period modulation, or quasi-biennal oscillation, of its magnetic activity as observed in the Sun and the 1-Gyr-old solar analog HD\,30495. In addition, the lithium abundance and the chromospheric activity estimated from \textsc{Hermes} confirms that KIC\,10644253 is a young and more active star than the Sun.
\end{abstract}



\section{Introduction}
The Mount Wilson spectroscopic observations of main-sequence G and K stars \citep{wilson78,duncan91} have suggested the existence of two distinct branches of cycling stars, the active and inactive \citep{saar92,soon93}, and that the Sun lies squarely between the two, thus appearing as a peculiar outlier \citep{bohm07}. Today, whether the solar dynamo and the related surface magnetic activity are typical or peculiar still remains an open question \citep{metcalfe16}. Finding solar-analog stars and studying their surface magnetic activity is a very promising way to understand solar variability and its associated dynamo \citep{egeland16}. Moreover, the study of the magnetic activity of solar analogs is also important for understanding the evolution of the Sun and its environment in relation to other stars and the habitability of their planets. 

The unprecedented quality of the continuous four-year photometric observations collected by the {\it Kepler} satellite \citep{borucki10} allowed the measurements of acoustic oscillations in hundreds of solar-like stars \citep{chaplin14}. Moreover, the length of the {\it Kepler} dataset provides a unique source of information for detecting magnetic activity and the associated temporal variability in the acoustic oscillations.
Indeed, it is well established that in the case of the Sun, p modes are sensitive to changes in the surface magnetic activity \citep{wood85}. Moreover, the p-mode frequencies are the only proxy that can reveal  inferences on sub-surface changes with activity that is not detectable at the surface by standard proxies \citep[e.g.,][]{salabert09,salabert15,basu12}.

\citet{cayrel96} provided a definition of a solar-analog star based on the fundamental parameters (e.g., $M$ and $T_\mathrm{eff}$). 
Here, we took advantage of the combination of asteroseismology with high-resolution spectroscopy 
which substantially improves the accuracy of the stellar parameters and reduces their errors \citep{mathur12,chaplin14,metcalfe14}. We selected stars for which solar-like oscillations were detected in order to avoid very active stars \citep{salabert03,mosser09,chaplin11}. This is what we called a seismic solar-analog star. We included in the sample only stars with measured rotation \citep{garcia14} to ensure the presence of magnetic activity.
A total of 18 seismic solar analogs were identified from the photometric {\it Kepler} observations \citep{salabert16a}.

\section{Photospheric and chromospheric magnetic activity of seismic solar analogs}
\begin{figure}[tbp]
\includegraphics[width=0.49\textwidth]{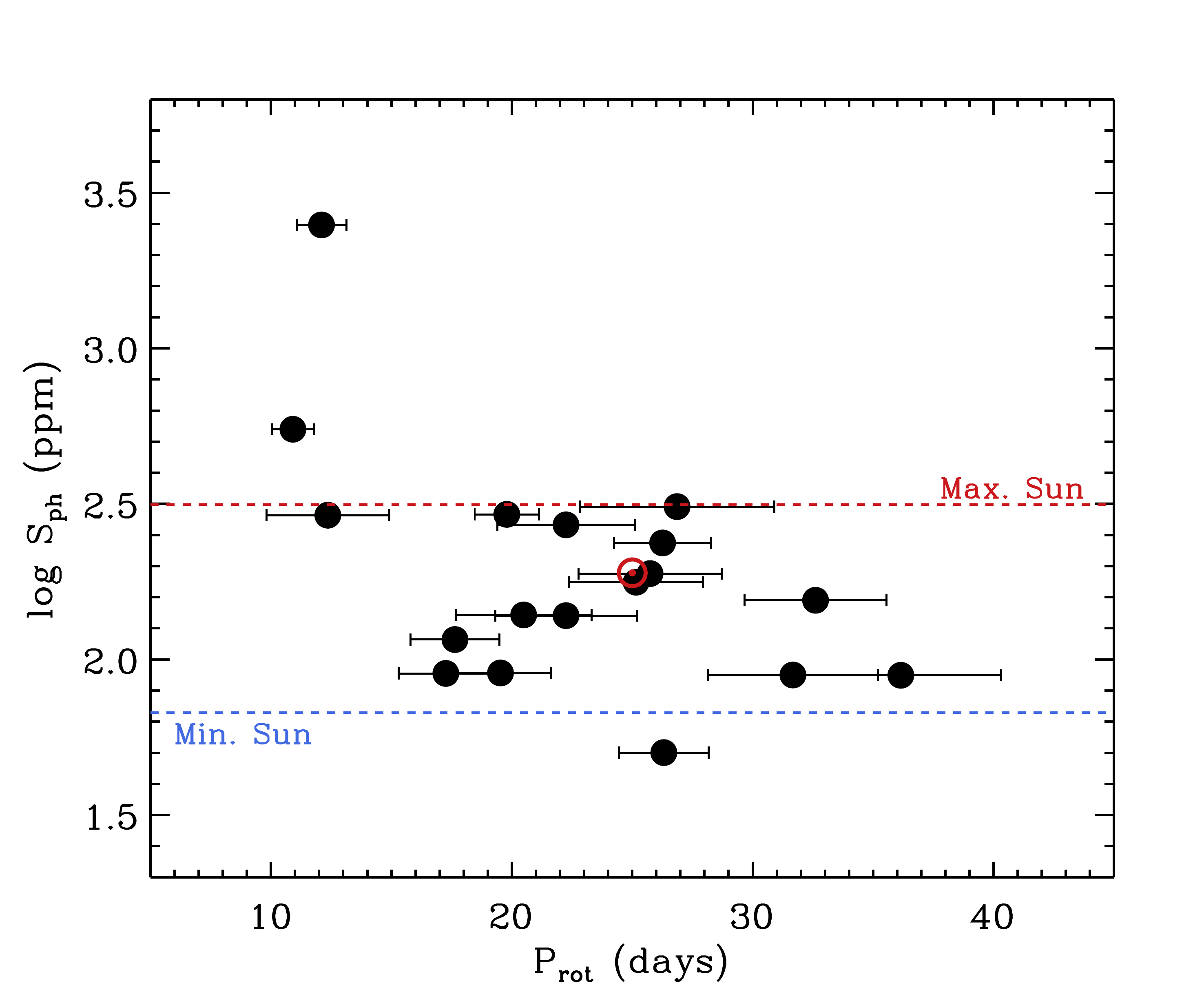}
\includegraphics[width=0.49\textwidth]{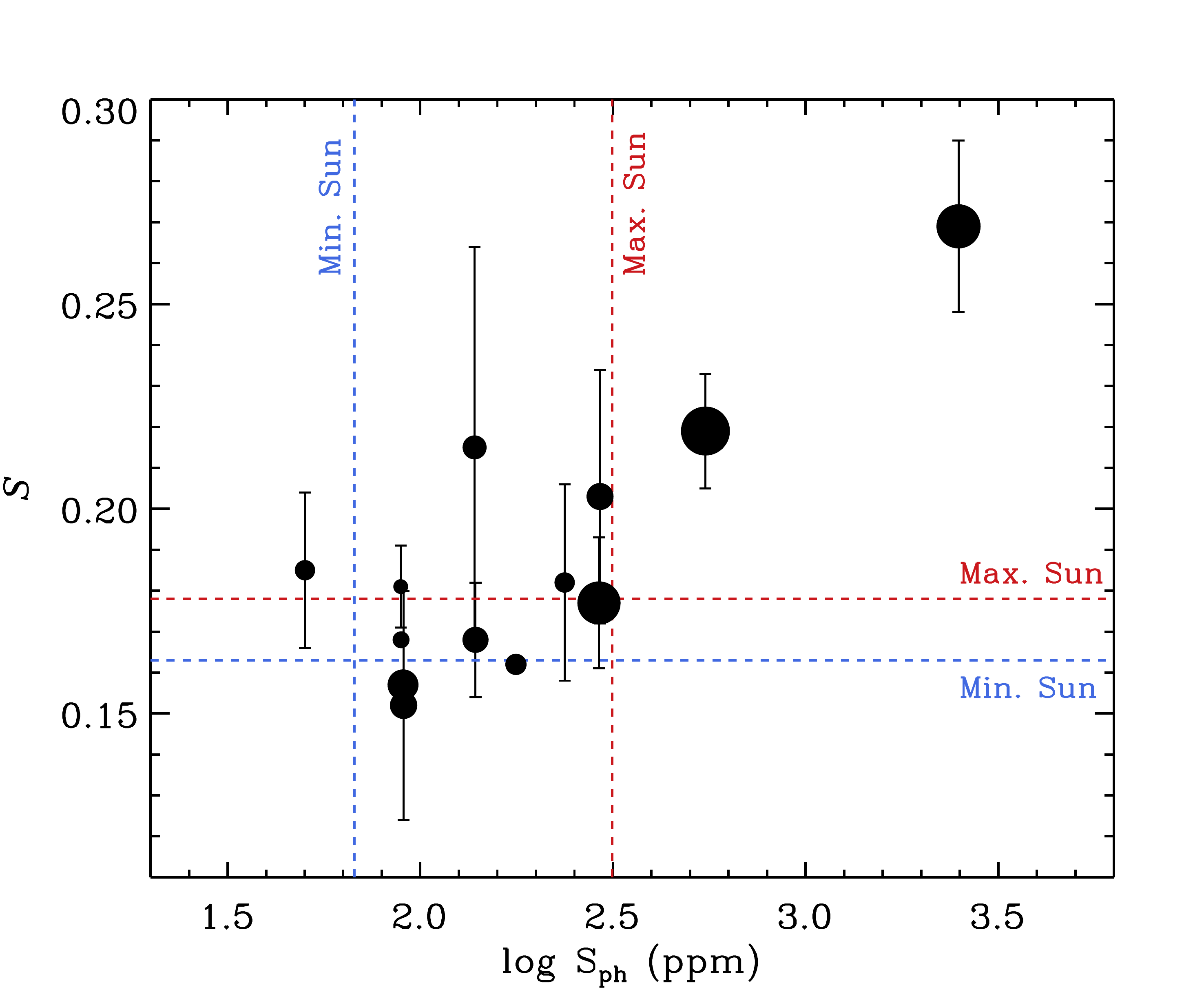}
\caption{(Left panel) Photospheric index $S_\mathrm{ph}$ as a function of the rotation period $P_\mathrm{rot}$ of the 18 seismic solar analogs observed with {\it Kepler}. (Right panel) Chromospheric $\mathcal{S}$~index derived from \textsc{Hermes} observations and calibrated into the MWO system as a function of the photospheric index $S_\textrm{ph}$. The symbol size is inversely proportional to the rotation. Adapted from \citet{salabert16a}.} \label{fig:fig1}  
\end{figure}

The photospheric activity $S_\textrm{ph}$ index corresponds to a proxy of the global stellar magnetic variability derived by means of the surface rotation $P_\mathrm{rot}$ \citep{mathur14b}. It is defined as the mean value of the light-curve fluctuations over sub series of length $5 \times P_\mathrm{rot}$. We note however that $S_\textrm{ph}$  represents a lower limit of the photospheric activity as it depends on the inclination angle. It was estimated here through the analysis of the {\it Kepler} long-cadence observations calibrated as described in \citet{garcia11}. The chromospheric activity $\mathcal{S}$ index is a proxy of the strength of the plasma emission in the cores of the Ca\textsc{ii}\,H\&K lines in the near ultra violet \citep{wilson78}. In this work, the $\mathcal{S}$ index was measured from spectroscopic observations collected with the \textsc{Hermes} spectrograph \citep{raskin11} mounted on the 1.2-m \textsc{Mercator} telescope at the Observatorio del Roque de los Muchachos (La Palma, Canary Islands, Spain). A detailed description of the data processing of the \textsc{Hermes} observations can be found in \citet{beck16a}.  We note that the result is dependent on the instrumental resolution and on the spectral type. However, as the selected stars were chosen for all having comparable stellar properties to the Sun, the estimated values of the $\mathcal{S}$~index can be thus safely compared between each other.

The left panel of Fig.~\ref{fig:fig1} shows the $S_\textrm{ph}$ of the 18 seismic solar analogs as a function of their rotation $P_\mathrm{rot}$. The mean value of the solar $S_\textrm{ph}$ over cycle~23 calculated from the photometric VIRGO/SPM observations \citep{frohlich95} is also indicated, as well as the corresponding values at minimum and maximum of activity. The $S_\textrm{ph}$ of the identified solar analogs is comparable to the Sun, within the range of activity covered over a solar cycle. Moreover, the two youngest solar analogs in our sample below 2\,Gyr-old \citep{chaplin14,metcalfe14},  which rotate in about 11~days, are the most active.
The comparison between the $S_\mathrm{ph}$ and the $\mathcal{S}$ magnetic activity proxies is shown on the right panel of Fig.~\ref{fig:fig1} for a subset of 13 stars with a S/N(Ca)\,$>$\,15 in the spectroscopic data \citep[for more explanations, see][]{salabert16a}. The values of $\mathcal{S}$ were calibrated to the Mount Wilson Observatory (MWO) system using the \textsc{Hermes} scaling factor derived by \citet{beck16a}. The mean values at minimum and maximum of solar activity are also represented. The resulting activity box corresponds to the range of change in solar activity along the 11-year magnetic cycle. Although the sample of stars is small, the $S_\mathrm{ph}$ and $\mathcal{S}$ indices are observed to be complementary, within the errors. We note also that both proxies were not estimated from contemporaneous {\it Kepler} and \textsc{Hermes} observations, introducing a dispersion partly related to possible temporal variations in stellar activity. Nevertheless, it confirms that $S_\mathrm{ph}$ can complement the classical $\mathcal{S}$~index for activity studies.

\section{Magnetic variability in the young solar analog KIC\,10644253}
With a rotation of $\sim$\,11~days, the solar analog KIC\,10644253 (BD+47\,2683, $V=9.26$) is the youngest solar-like pulsating star observed by {\it Kepler} with an age of $1.07\,\pm\,0.25$\,Gyr \citep{metcalfe14} and one of the most active \citep{garcia14}. It is thus an excellent candidate for investigating the magnetic activity of a young Sun with asteroseismic data. 
In addition to the Sun, temporal variations of p-mode frequencies related to magnetic activity were so far observed in only three stars: the F-type stars HD\,49933 \citep{garcia10} and KIC\,3733735 \citep{regulo16}, and the solar-analog G-type KIC\,10644253 \citep{salabert16b}. 

\begin{figure*}[tbp]
\floatbox[{\capbeside\thisfloatsetup{capbesideposition={right,bottom},capbesidewidth=5.4cm}}]{figure}[\FBwidth]
{\includegraphics[width=0.65\textwidth,angle=0]{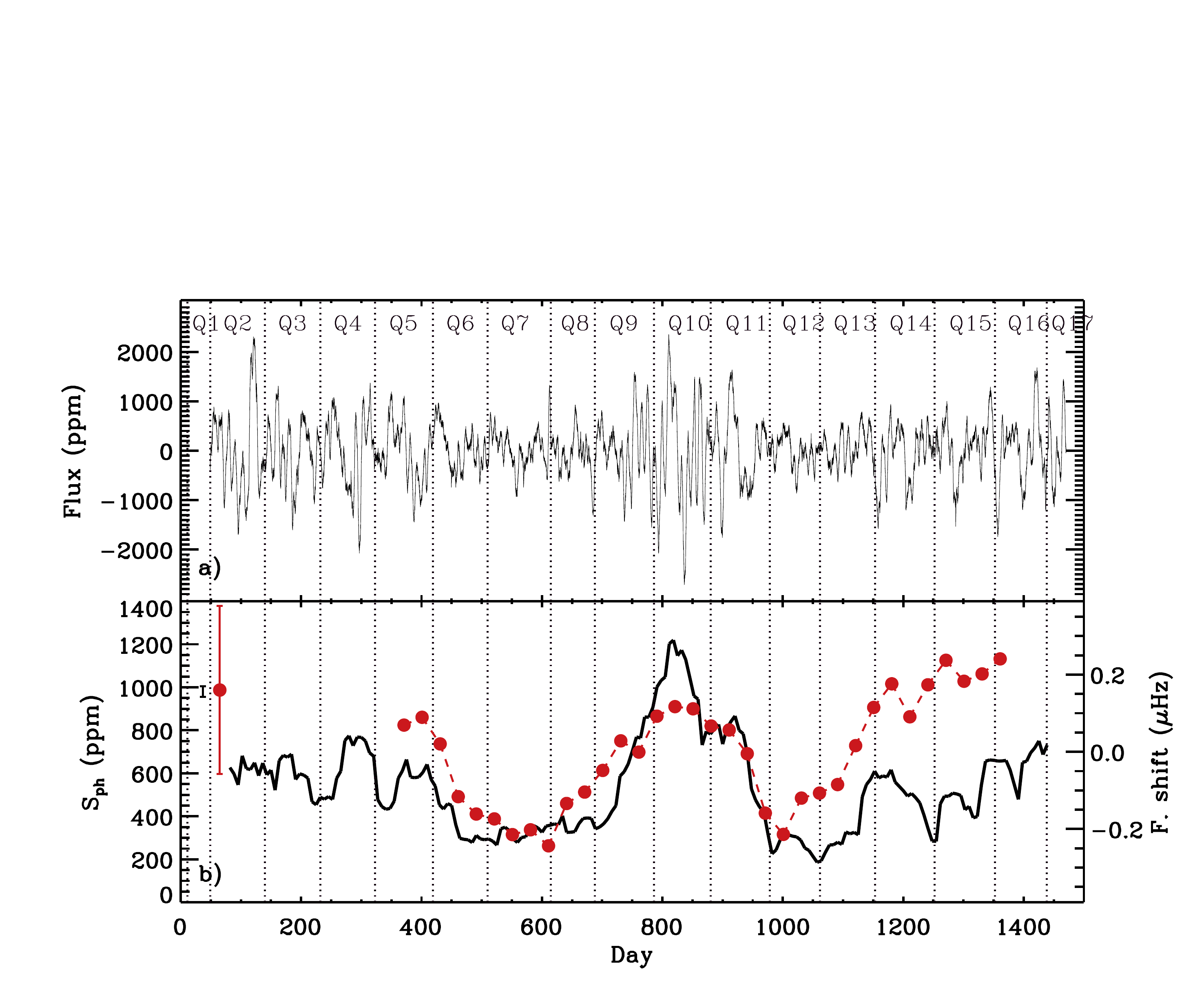}}
{\caption{\label{fig:fig2}  (Top panel) Photometric long-cadence observations of KIC\,10644253 collected over 1411~days by {\it Kepler} as a function of time. 
(Bottom panel) Photospheric index $S_{\textrm{ph}}$ (black) of KIC\,10644253 as a function of time compared to the frequency shifts obtained from the cross-correlation analysis (red circles). The frequency shifts were extracted from the continuous short-cadence observations from Q5 to Q17. The associated mean uncertainties are illustrated in the upper left-hand corner using the same color code. In the two panels, the vertical dotted lines represent the observational length of each {\it Kepler} quarter from Q1 to Q17. Adapted from \citet{salabert16b}.}}
\end{figure*} 

To study the temporal variations of the low-degree, p-mode oscillation frequencies observed in KIC\,10644253, the {\it Kepler}  short-cadence dataset was split into contiguous 180-day-long sub series \citep{salabert16b}. The associated power spectra were analyzed using both peak-fitting \citep{ballot11,salabert11} and cross-correlation \citep{regulo16} independent methods and the corresponding frequency shifts $\langle  \delta \nu\rangle$ extracted. In addition, the light curve was analyzed to estimate the $S_\mathrm{ph}$ over sub series of $5\,\times\, P_\mathrm{rot}\,=\,54.55$\, days.
Figure~\ref{fig:fig2} shows that both the photospheric $S_\mathrm{ph}$ and the frequency shifts $\langle \delta\nu \rangle$ present the signature of magnetic activity variability. A modulation of about 1.5\,years  is measured in both observables of about 900\,ppm for $S_\mathrm{ph}$ and 0.5\,$\mu$Hz for the frequency shifts. It could be the signature of the short-period modulation, or quasi-biennal oscillation, of its magnetic activity as observed in the Sun \citep[see, e.g.,][]{fletcher10}. The variations found in KIC\,10644253 at a rotation period of $\sim$\,11~days is analogous to what is found by \citet{egeland15} from the study of the temporal variations of the $\mathcal{S}$ index in the solar analog HD\,30495 falling on the inactive branch \citep{bohm07}. Moreover, the comparison between magnitude and frequency dependence of the
frequency shifts measured for KIC\,10644253 with the ones obtained for the Sun indicates that the same physical mechanisms are involved in the sub-surface layers in both stars. 

In addition, the analysis of the \textsc{Hermes} spectroscopic observations shows that KIC\,10644253 is $\sim$\,18\% chromospherically more active than the Sun with an $\mathcal{S}$ index of $0.213\pm0.008$. Moreover, the high lithium abundance of $2.74\pm0.03$\,dex and the effective temperature of $6006\pm100$\,K  mean that the lithium at the surface has not been depleted yet by internal processes \citep{ramirez12}. This is validating its young age estimated from seismology and in agreement with a rotation of $\sim$\,11~days from gyrochronology \citep{meibom11,jen16}. Furthermore, among the 18 solar analogs in this sample, KIC\,10644253 has the highest lithium abundance (Beck et al., Submitted).

\section{Conclusions}
The study of the characteristics of the surface activity of solar analogs can provide new constraints in order to better understand the magnetic variability of the Sun, and its underlying dynamo during its evolution.
We analyzed here the sample of main-sequence stars observed by the {\it Kepler} satellite for which solar-like oscillations were detected and rotational periods measured and from published stellar parameters, we identified 18 seismic solar analogs.
We then studied the properties of the photospheric and chromospheric magnetic activity of these stars in relation of the Sun. The photospheric index $S_\mathrm{ph}$ was derived through the analysis of the {\it Kepler} observations, while the chromospheric proxy $\mathcal{S}$ was measured with follow-up, ground-based \textsc{Hermes} spectroscopic observations. We showed that the magnetic activity of the Sun is comparable to the activity of the seismic solar analogs studied here, within the maximum-to-minimum activity variations of the Sun during the 11-year cycle. As expected, the youngest and fastest rotating stars are observed to be the most active of our sample. Furthermore, the comparison of the photospheric index $S_\mathrm{ph}$ with the well-established chromospheric $\mathcal{S}$~index shows that $S_\mathrm{ph}$ can be used to provide a suitable magnetic activity proxy. We established the existence of a temporal variability of the magnetic activity was observed in the young (1 Gyr-old) solar analog KIC\,10644253. A significant modulation of about 1.5 years was measured in the low-degree, p-mode frequencies and in the photospheric index $S_\mathrm{ph}$. 



\begin{acknowledgements}
The rest of the Seismic Solar Analogs (SSA) team consists by alphabetic order of:
J. Ballot, L. Bigot, O.~L. Creevey, R. Egeland, S. Mathur, T.~S. Metcalfe, J.-D. do Nascimento Jr., P.~L. Pall\'e, F. P\'erez Hern\'andez, and C. R\'egulo.
The authors wish to thank the entire {\it Kepler} team. Funding for this Discovery mission is provided by NASA Science Mission Directorate. The spectroscopic observations are made with the Mercator Telescope, operated by the Flemish Community at the Spanish Observatorio del Roque de los Muchachos (La Palma) of the Instituto de Astrof\'isica de Canarias. This work utilizes data from the National Solar Observatory/Sacramento Peak Monitoring Program, managed by the National Solar Observatory, which is operated by the Association of Universities for Research in Astronomy (AURA), Inc. under a cooperative agreement with the National Science Foundation. The research leading to these results has received funding from the European Community's Seventh Framework Program ([FP7/2007-2013]) under grant agreement no. 312844 (SPACEINN). DS and RAG acknowledge the financial support from the CNES GOLF and PLATO grants.  
PGB acknowledges the ANR (Agence Nationale de la Recherche, France) program IDEE (no. ANR-12-BS05-0008) "Interaction Des Etoiles et des Exoplan\`etes".
\end{acknowledgements}



%
\end{document}